\documentclass{article}
\usepackage{spconf,amsmath,graphicx}
\usepackage{booktabs}
\usepackage{tabularx}

\title{T-GSA: Transformer with Gaussian-Weighted Self-Attention  
\\ for Speech Enhancement}
\twoauthors
  {Jaeyoung Kim}
   {jykim1@alumni.stanford.edu}
  { Mostafa El-Khamy, Jungwon Lee}
   {SOC R\&D, Samsung Semiconductor, Inc. USA\\ 
   Emails:\{mostafa.e, jungwon2.lee\}@samsung.com}

\begin{document}
%
\maketitle
\begin{abstract}
Transformer neural networks (TNN) demonstrated state-of-art performance  on many natural language processing (NLP) tasks, replacing recurrent neural networks (RNNs), such as LSTMs or GRUs.  However, TNNs did not perform well in speech enhancement, whose contextual nature is different than NLP tasks, like machine translation. Self-attention is a core building block of the Transformer, which not only enables parallelization of sequence computation, but also provides the constant path length between symbols that is essential to learning long-range dependencies.
In this paper, we propose a  Transformer with Gaussian-weighted self-attention (T-GSA), whose attention weights are attenuated according to the distance between target and context symbols. The experimental results show that the proposed T-GSA has significantly improved speech-enhancement performance, compared to the Transformer and RNNs. 
\end{abstract}

\section{Introduction}
\label{sec:intro}

Deep neural networks have shown great success in speech enhancement~\cite{narayanan2013ideal,erdogan2015phase,wang2014training,pascual2017segan,rethage2018wavenet, soni2018time} and performed better than the popular model-based statistical approaches, such as MMSE STSA~\cite{ephraim1984speech} or OM-LSA~\cite{ephraim1985speech, cohen2001speech}.

 Recurrent neural networks (RNNs), such as LSTM~\cite{hochreiter1997long} or GRU~\cite{chung2014empirical} were the most popular neural network architectures in speech enhancement, due to their powerful sequence learning. Recently, the Transformer~\cite{vaswani2017attention} was presented as a new sequence-learning architecture with significant improvements over RNNs in machine translation and many other natural language processing tasks. The Transformer uses a self-attention mechanism to compute symbol-by-symbol correlations in parallel, over the entire input sequence, which are used to predict the similarity between the target and neighboring context symbols. The predicted similarity vector is normalized by the softmax function and used as attention weights to combine context symbols. 

  Unlike RNNs, the Transformer can process an input sequence in parallel, which can significantly reduce training and inference times. Moreover, the Transformer provides a fixed path length, that is the number of time steps to traverse, before computing attention weights or symbol correlations. Typically, RNNs have the path length proportional to the distance between target and context symbols due to sequential processing, which makes it difficult to learn long-range dependencies between symbols. The Transformer resolved this issue with the self-attention mechanism. 
  
Current Transformer networks did not show improvements in  acoustic signal processing, such as speech enhancement or speech denoising. The fixed path length property that benefited many NLP tasks is not compatible with the physical characteristics of acoustic signals, which tend to be more correlated with the closer components. Therefore, positional encoding is required to penalize attention weights according to the acoustic signal characteristics, such that less attention is provided to more distant symbols. In this paper, we propose a Transformer with Gaussian-weighted self-attention (T-GSA), whose attention weights are attenuated according to the distance between correlated symbols. The attenuation is determined by the Gaussian variance which can be learned during training. Our evaluation results show that the proposed T-GSA significantly improves over existing Transformer architectures, as well as over the former best recurrent model based on the LSTM architecture.

\section{Proposed Architectures}
\label{sec:proposed}
\begin{figure}[htb]
  \centering
  \centerline{\includegraphics[width=4.5cm]{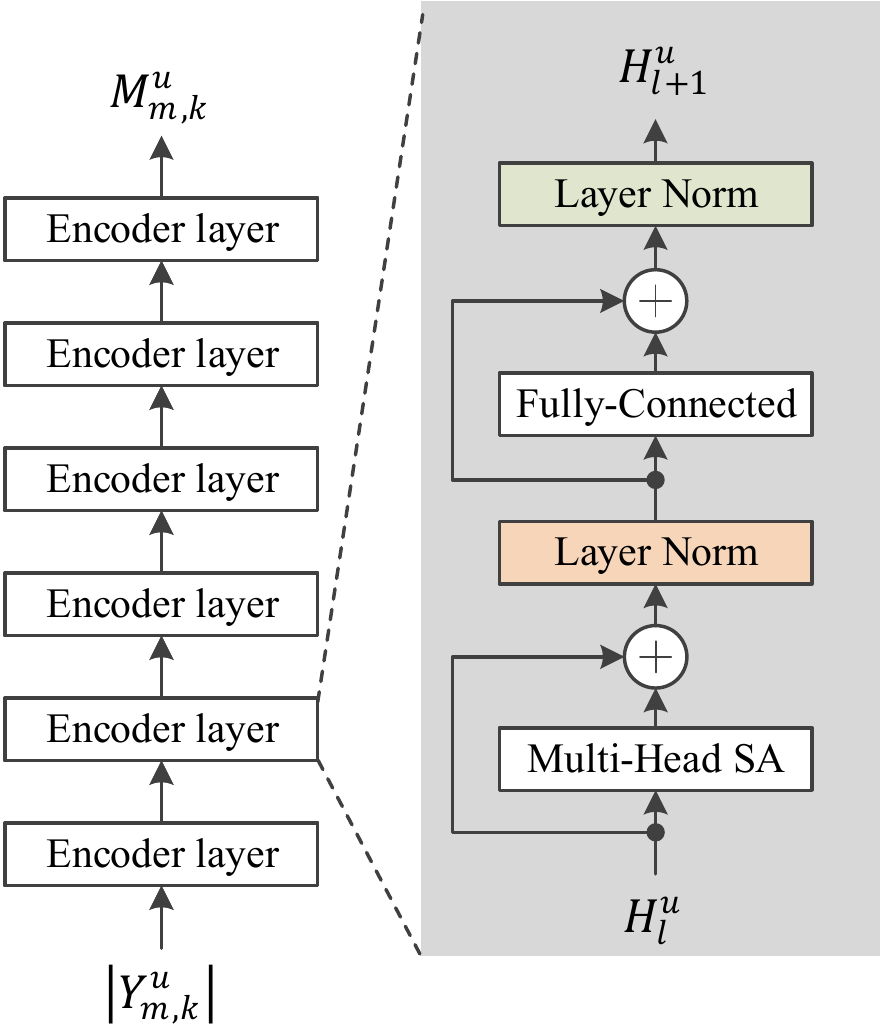}}
\caption{Block diagram of the Transformer encoder for speech enhancement}
\label{fig:real_tf}
\end{figure}
Figure~\ref{fig:real_tf} shows the proposed denoising network based on the Transformer encoder architecture. The original Transformer consists of encoder and decoder networks. In speech denoising, the input and output sequences have the same length. 
Hence, we only used the encoder network and alignment between input and output sequences is not necessary.  The network input, $|Y^u_{m,k}|$, is the short-time Fourier transform (STFT) spectrum magnitude of the  noisy time-domain speech $y^u(n)$. $u$ is the  utterance index, $m$ is the frame index, and $k$ is the frequency index. The input noisy signal is given by
\begin{equation}
y^u(n)=x^u(n)+n^u(n),
\end{equation}
where $x^u(n)$ and $n^u(n)$ are the clean and noisy speech  signals, respectively. Each encoder layer consists of multi-head self-attention, layer normalization and fully-connected layers, which is the same as the original Transformer encoder. The network output is a time-frequency mask that predicts clean speech by scaling the noisy input: 
\begin{equation}
|\hat{X}^u_{m,k}|=M^u_{m,k}|Y^u_{m,k}|.
\end{equation}
The estimated clean  spectrum magnitude $|\hat{X}^u_{m,k}|$ is multiplied with the phase of the input spectrum, from which the time-domain signal, $\hat{x}^u(n)$, is obtained by the inverse short-time Fourier transform (ISTFT).

\subsection{GSA: Gaussian-weighted Self-Attention}
\begin{figure}[htb]
  \centering
  \centerline{\includegraphics[width=7cm]{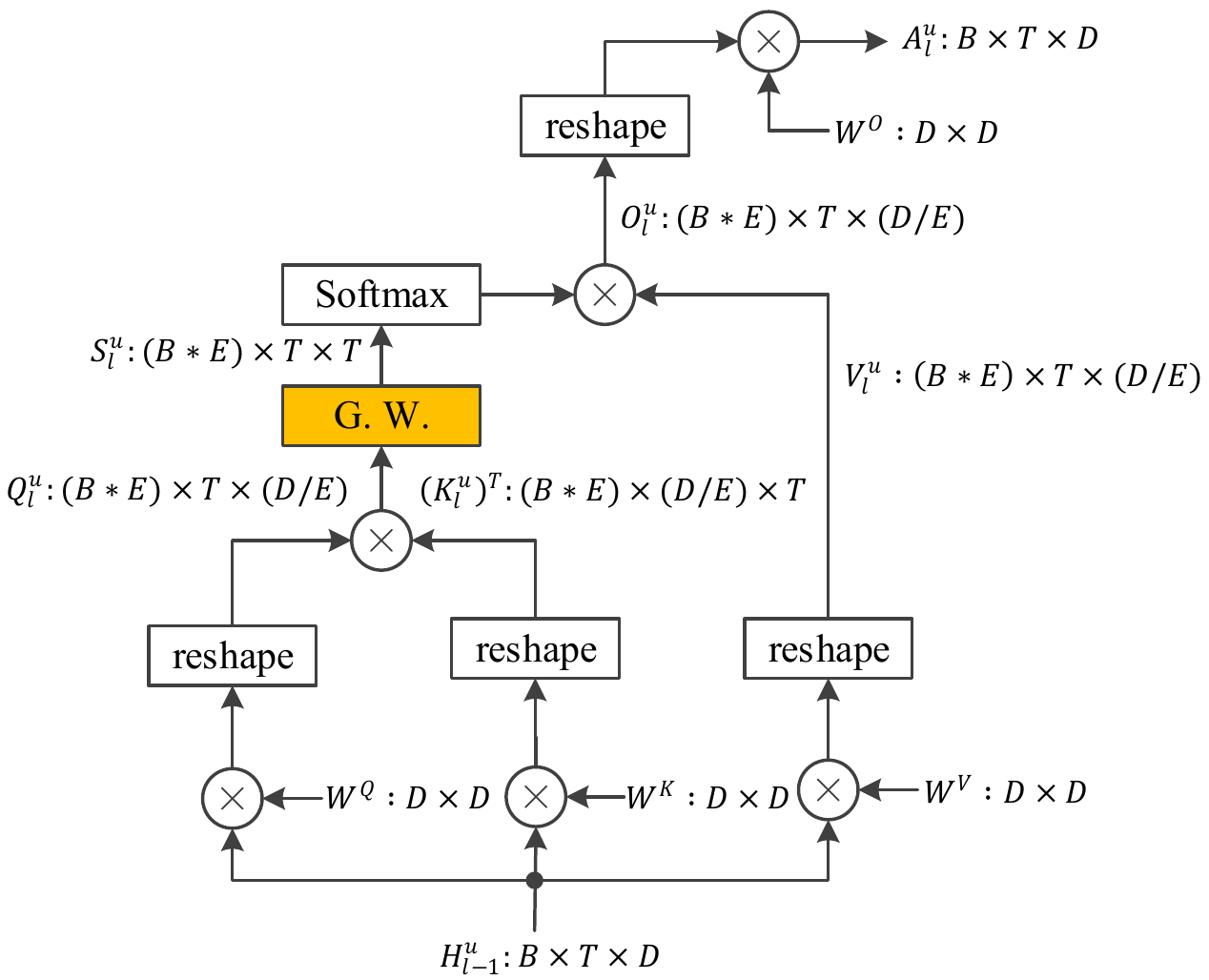}}
\caption{Block diagram of the proposed multi-head self-attention: The G.W. block performs element-wise multiplication of the Gaussian-weight matrix with the generated score matrix. The matrix dimensions are noted besides each signal. }
\label{fig:gw}
\end{figure}
Figure~\ref{fig:gw} describes the proposed Gaussian-weighted multi-head self-attention. $B$, $T$ and $D$ are the batch size, sequence length, and input dimension. $E$ is the number of self-attention units. Query, key and value matrices are defined as follows:
\begin{eqnarray}
\label{eq:dm3}
Q^u_l&=&W^QH_{l-1}^u  \\  
K^u_l&=&W^KH_{l-1}^u  \\  
V^u_l&=&W^VH_{l-1}^u 
\end{eqnarray}
where $H^u_{l}$ is $l^{th}$ hidden layer output. $W^Q$, $W^K$, and $W^V$ are network parameters.

The multi-head attention module in our proposed T-GSA is modified by deploying a Gaussian weighting matrix to scale the score matrix, which is computed from the key and query matrix multiplication as follows:
\begin{equation}
\label{eq:sl}
S^u_l=G_l\circ\left( \frac{Q^u_l(K^u_l)^T}{\sqrt{d}}\right) = G_l\circ C_l^u
\end{equation} 
$G_l$ is the Gaussian weight matrix which is element-wise multiplied with the score matrix, $C_l^u$. The proposed Gaussian weighting matrix is calculated as follows:
\begin{equation}
G_l=
\left[
\begin{matrix}
g^l_{1,1} & g^l_{1,2} &\cdots &  g^l_{1,T}\\
g^l_{2,1} & g^l_{2,2} &\cdots &  g^l_{2,T}\\
        &  \vdots       &       &          \\ 
g^l_{T,1} & g^l_{T,2} &\cdots &  g^l_{T,T}\\
\end{matrix}
\right]
\end{equation}
where $g^l_{i,j}$ is $e^{\frac{-|i-j|^2}{\sigma_l^2}}$, $i$ is a target frame index, $j$ is a context frame index and $\sigma_l$ is a trainable parameter that determines the weight variance. For example, $g^l_{i,j}$ corresponds to the scaling factor for the context frame $j$ when the target frame index is $i$. The diagonal terms in $G_l$ correspond to the scaling factors for the target frames, which is always set to be 1. $g^l_{i,j}$ is inversely proportional to the distance between the target and context frames to provide larger attenuation of the attention given to the more distant context frames and smaller attenuation for the closer ones. Since we let $\sigma_l$ to be a trainable parameter, context localization can be learned by the acoustic training data consisting of clean and noisy speech signals.
After the softmax function, the self-attention matrix is multiplied by the value matrix $V^u_l$:
\begin{equation}
O_l^u=\textrm{SoftMax}\left(|S^u_l | \right)V^u_l
\end{equation}
One thing to note is that the absolute value of the matrix $S^u_l$ is applied to the softmax function. The reason for this is that unlike NLP tasks, the negative correlation information in the signal estimation is as important as the positive correlation. By taking the absolute value of the Gaussian weighted score matrix, the resultant self-attention weights will only depend on the score magnitude, which enables to equally utilize both positive and negative correlations.

\emph{Remark 1:} The attention biasing~\cite{sperber2018self} used for an acoustic model design is a different positional encoding scheme, where additive bias is applied to the score matrix in the self-attention block. Different from our proposed GSA, the additive bias can totally alter the attention signs which depend on whether there is positive or negative correlation with the symbol which is attended to. However, our proposed GSA preserves the correlation sign, and just alters its scale according to the distance from the attended symbol.
%

\subsection{Extension to Complex Transformer Architecture}
\label{sec:complex}
\begin{figure}[htb]
  \centering
  \centerline{\includegraphics[width=6cm]{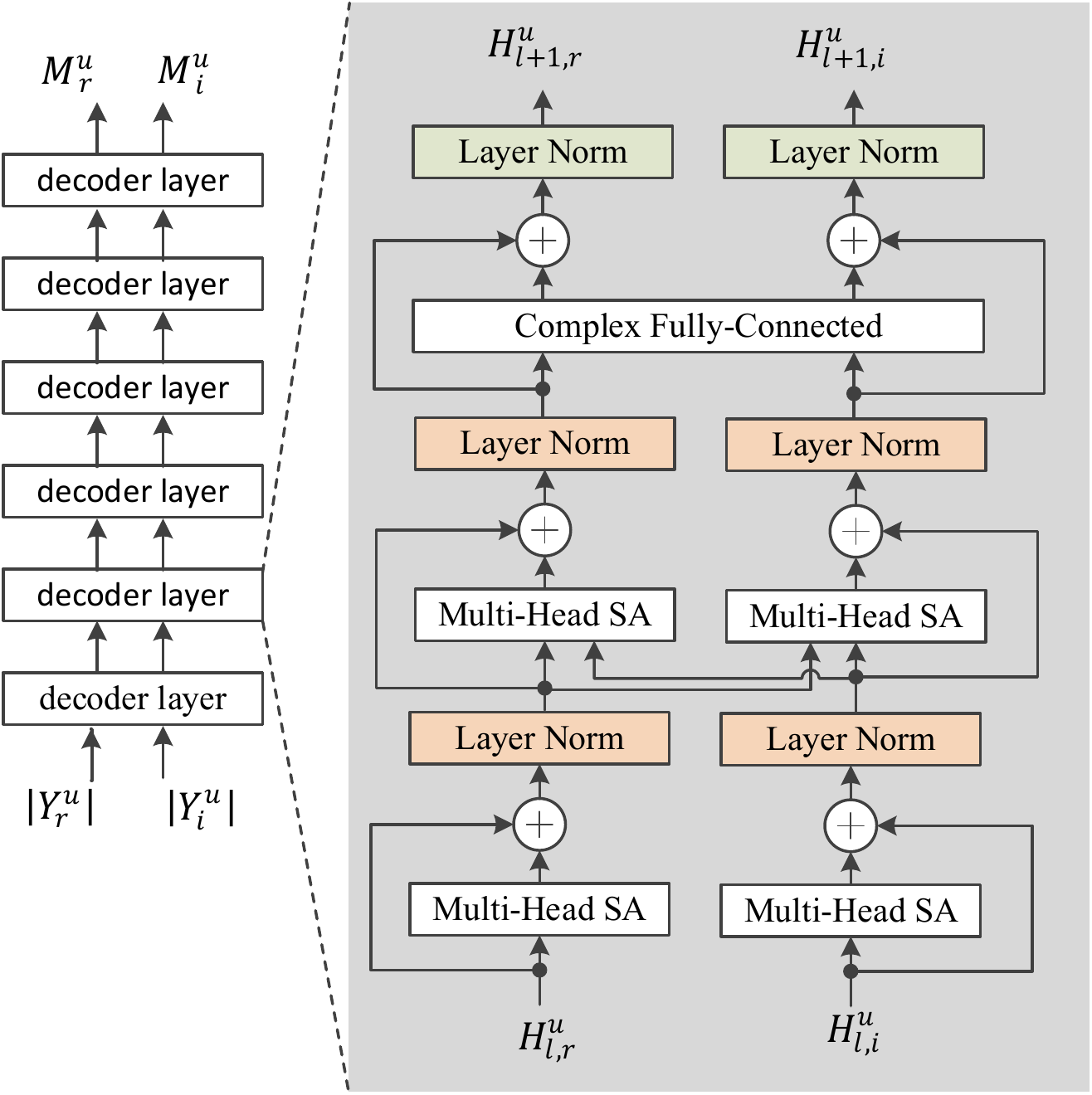}}
\caption{Block Diagram of Complex Transformer architecture}
\label{fig:cplx}
\end{figure}
We proposed a complex Transformer architecture for speech enhancement, as shown in Figure~\ref{fig:cplx}. Compared with the real Transformer architecture in Figure~\ref{fig:real_tf}, the complex Transformer has two inputs and two outputs, corresponding to the real and imaginary parts of the input and output STFTs, respectively. The network inputs, $Y^u_r$ and $Y^u_i$, are the real and imaginary parts of the input noisy spectrum. By estimating both the real and complex parts of the output clean speech spectrum, the complex Transformer denoiser showed significantly better SDR and PESQ performance. The network output is a complex mask that generates the complex denoised output $\hat{X}^u_{m,k}$ as follows:
\begin{eqnarray}
\label{eq:dm3}
\hat{X}^u_{r,m,k}&=& |Y^u_{r,m,k}| M^u_{r,m,k} - |Y^u_{i,m,k}|M^u_{i,m,k}  \\  
\hat{X}^u_{i,m,k}&=& |Y^u_{r,m,k}| M^u_{i,m,k} + |Y^u_{i,m,k}|M^u_{r,m,k}
\end{eqnarray}
where a subscript $r$ means a real part and a subscript $i$ corresponds to an imaginary part.
The right grey block in Figure~\ref{fig:cplx} describes the decoder layer of the complex Transformer network. $H^u_{l,r}$ and $H^u_{l,i}$ are the real and imaginary outputs of the $l^{th}$ layer, respectively. The first multi-head self attention blocks are applied to each real and imaginary input separately. After layer normalization, the second multi-head attention gets mixed inputs from the real and imaginary paths. For example, the left second multi-head attention gets the right-side layer normalization output as key and value input in Figure~\ref{fig:gw}. The query input comes from the left-side layer normalization. The main idea is to exploit the cross-correlation between the real and imaginary parts by mixing them in the attention block. After another layer normalization, a complex fully-connected layer is applied. The complex fully-connected layer has real and imaginary weights and the standard complex operation is performed on the complex input from the second layer normalization.

\subsection{End-to-End Metric Optimization}
A multi-task denoising scheme has been recently proposed to train speech enhancement networks by jointly optimizing both the Signal to Distortion Ratio (SDR) and the Perceptual Evaluation of Speech Quality (PESQ) metrics \cite{kim2019end}. The proposed denoising framework outperformed the existing spectral mask estimation schemes~\cite{erdogan2015phase,narayanan2013ideal, wang2014training} and generative models~\cite{pascual2017segan, rethage2018wavenet, soni2018time} to provide a new state of the art performance. We adopt the overall training framework shown in Figure 2 in~\cite{kim2019end} to train our networks. First, the denoised complex spectrum is transformed into the time-domain acoustic signal via Griffin-Lim ISTFT~\cite{griffin1984signal}. Second, the proposed SDR and PESQ loss functions in~\cite{kim2019end} are computed based on the acoustic signal. The two loss functions are jointly optimized by this combined loss function:  
 \begin{equation}
\textrm{L}_{\textrm{SDR-PESQ}}=\textrm{L}_{\textrm{SDR}}+\alpha \textrm{L}_{\textrm{PESQ}},
\end{equation}
where $\textrm{L}_{\textrm{SDR}}$ and $\textrm{L}_{\textrm{PESQ}}$ are SDR and PESQ loss functions defined in Eq. 20 and 32 in~\cite{kim2019end}, respectively. $\alpha$ is a hyper-parameter to adjust relative importance between the SDR and PESQ loss functions, and is set to be $3.2$ after grid-search on the validation set. 

\section{Experimental Results}
\label{sec:exp}
\begin{table*}[ht]
\caption{SDR and PESQ results on QUT-NOISE-TIMIT: Test set consists of 6 SNR ranges:-10, -5, 0, 5, 10, 15 dB. The highest SDR or PESQ scores for each SNR test data were highlighted with bold fonts.}
\label{tab:tq}
\vskip 0.1in
\centering
\begin{tabularx}{\textwidth}{l r r r r r r | r r r r r r }
\toprule
	& \multicolumn{6}{c|}{SDR}	& \multicolumn{6}{c}{PESQ} \\
Loss Type	& -10 dB & -5 dB & 0 dB & 5 dB & 10 db & 15 dB & -10 dB & -5 dB & 0 dB & 5 dB & 10 db & 15 dB\\
\midrule
Noisy Input & -11.82 & -7.33 & -3.27 & 0.21 & 2.55 & 5.03 & 1.07 & 1.08 & 1.13 & 1.26 & 1.44 & 1.72 \\
CNN-LSTM & -2.31 & 1.80 & 4.36 & 6.51 & 7.79 & 9.65 & 1.43 & 1.65 & 1.89 & 2.16 & 2.35 & 2.54   \\  
O-T & -3.25 & 0.92 & 3.39 & 5.35 & 6.39 & 8.10 & 1.29 & 1.45 & 1.63 & 1.87 & 2.07 & 2.29   \\  
T-AB & -2.80 & 1.18 & 3.67 & 5.67 & 6.78 & 8.18 & 1.49 & 1.67 & 1.85 & 2.01 & 2.28 & 2.50   \\  
T-GSA (ours) & -1.66 & 2.35 & 4.95 & 7.10 & 8.40 & 10.36 & \textbf{1.54} & \textbf{1.76} & \textbf{2.00} & \textbf{2.28} & \textbf{2.51} & \textbf{2.74} \\
C-T-GSA (ours) & \textbf{-1.57} & \textbf{2.51} & \textbf{5.03} & \textbf{7.36} & \textbf{8.58} & \textbf{10.40} & 1.43 & 1.64 & 1.88 & 2.17 & 2.40 & 2.67\\ 

\bottomrule
\end{tabularx}
\end{table*}

\subsection{Experimental Settings}
Two datasets were used for training and evaluation of the proposed Transformer architectures:

\textbf{\underline{QUT-NOISE-TIMIT}}~\cite{dean2010qut}: QUT-NOISE-TIMIT is synthesized by mixing 5 different background noise sources with the TIMIT~\cite{garofolo1993darpa}. For the training set, -5 and 5 dB SNR data were used but the evaluation set contains all SNR ranges. The total length of train and test data corresponds to 25 hours and 12 hours, respectively. The detailed data selection is described at Table 1 in~\cite{kim2019end}.  

\textbf{\underline{VoiceBank-DEMAND}}~\cite{valentini2016investigating}: 30 speakers selected from Voice Bank corpus~\cite{veaux2013voice} were mixed with 10 noise types: 8 from Demand dataset~\cite{thiemann2013diverse} and 2 artificially generated one. Test set is generated with 5 noise types from Demand that does not coincide with those for training data.

\subsection{Main Result}
\label{sec:main}

Table~\ref{tab:tq} shows SDR and PESQ performance of Transformer models on the QUT-NOISE-TIMIT corpus. CNN-LSTM is the prior best performing recurrent model which is comprised of convolutional and LSTM layers. Its network architecture is described at Section 3 in ~\cite{kim2019end}. O-T represents the original Transformer encoder, T-AB is the Transformer model with attention biasing explained in Remark 1, T-GSA is the real Transformer with Gaussian-weighted self-attention, and C-T-GSA is the complex Transformer model. The real transformers consisted of 10 encoder layers and the complex Transformer has 6 decoder layers. The encoder and decoder layers were described in Figure~\ref{fig:real_tf} and~\ref{fig:cplx} and they have 1024 input and output dimensions.  All the neural network models evaluated in this section were trained to minimize $\textrm{L}_{\textrm{SDR-PESQ}}$.

First, O-T showed large performance degradation compared with CNN-LSTM over all SNR ranges. Second, the T-AB substantially improved SDR and PESQ performance over O-T, which suggested that the positional encoding is an important factor to improve Transformer performance on this denoising problem. However, the T-AB still suffered from the large loss compared with the recurrent model, CNN-LSTM. Finally, with the proposed Gaussian-weighting, the T-GSA model significantly outperformed all the previous networks including CNN-LSTM. Especially, the large performance gap between attention biasing and Gaussian weighting suggested that using negative correlations is as important as using positive ones. 

The complex Transformer showed 0.1 to 0.2 dB SDR improvement over all the SNR ranges compared with the real Transformer. However, the PESQ performance degraded compared with the real Transformer. The reason for degradation could be overfitting due to the larger parameter size or due to the difficulty in predicting the phase spectrum. We are considering future research to make the complex network provide consistent performance gains on both the SDR and PESQ metrics.

\subsection{Comparison with Generative Models}

\begin{table}[ht]
\caption{Evaluation on VoiceBank-DEMAND corpus}
\label{tab:ge}
\vskip 0.1in
\centering
\begin{tabularx}{\columnwidth}{l X X X X X X}
\toprule
Models	& CSIG	& CBAK	& COVL	& PESQ	& SSNR & SDR\\
\midrule
Noisy Input	&	3.37	&	2.49	&	2.66	&	1.99	&	2.17 & 8.68\\
SEGAN		&	3.48	&	2.94	&	2.80	&	2.16	&	7.73 & -\\
WAVENET		&	3.62	&	3.23	&	2.98	&	-		&	-	& - \\
TF-GAN		&	3.80	&	3.12	&	3.14	&	2.53	&	- & - \\
CNN-LSTM	&	4.09	&	3.54	&	3.55	&	3.01	&	10.44 & 19.14\\
T-GSA	(ours) &	\textbf{4.18}	&	\textbf{3.59}	&	\textbf{3.62}	&	\textbf{3.06}	&	\textbf{10.78} & \textbf{19.57} \\

\bottomrule
\end{tabularx}
\end{table}

Table~\ref{tab:ge} shows comparisons with other generative models. All the results except CNN-LSTM and T-GSA (SEGAN~\cite{pascual2017segan}, WAVENET~\cite{rethage2018wavenet} and TF-GAN~\cite{soni2018time}) are from the original papers. CSIG, CBAK and COVL are objective measures where high value means better quality of speech~\cite{hu2008evaluation}. CSIG is mean opinion score (MOS) of signal distortion, CBAK is MOS of background noise intrusiveness and COVL is MOS of the overall effect. SSNR is Segmental SNR defined in~\cite{quackenbush1986objective}. 

The proposed Transformer model outperformed all the generative models for all the perceptual speech metrics listed in Table~\ref{tab:ge} with large margin. The main improvement came from the joint SDR and PESQ optimization schemes in~\cite{kim2019end} that benefited both CNN-LSTM and T-GSA. Furthermore, T-GSA showed consistently better performance over CNN-LSTM for all the metrics, which agrees with the result at Table~\ref{tab:tq}.
\section{Conclusion}
\label{sec:related}
We proposed a Transformer architecture with Gaussian-weighted self-attention for speech enhancement. The attention weights are attenuated proportionally to the distance between the target frame and the symbols attended to, while preserving the correlation signs. The performance evaluation result showed that the proposed self-attention scheme significantly improved both the SDR and PESQ scores over previous state-of-art recurrent and transformer networks. 

\bibliographystyle{IEEEbib}
\bibliography{strings}

\end{document}